\documentclass[sigconf]{acmart}

\AtBeginDocument{%
  }

\setcopyright{cc}
\setcctype{by}
\copyrightyear{2026}
\acmYear{2026}
\acmDOI{3805712.3808467}
\acmConference[SIGIR '26]{Proceedings of the 49th International ACM SIGIR Conference on Research and Development in Information Retrieval}{July 20--24, 2026}{Melbourne, VIC, Australia}
\acmBooktitle{Proceedings of the 49th International ACM SIGIR Conference on Research and Development in Information Retrieval (SIGIR '26), July 20--24, 2026, Melbourne, VIC, Australia}
\acmISBN{979-8-4007-2599-9/2026/07}

\usepackage{multirow}
\usepackage{booktabs}
\usepackage{enumitem}
\usepackage{siunitx}
\usepackage{makecell}
\usepackage{dsfont}
\usepackage{threeparttable} 
\usepackage{svg}
\usepackage{balance}

\begin{document}

\title[Don’t Contrast the Impossible: Region-Constrained Batching for Contrastive \\ User Modeling on a Local Community Platform]{Don’t Contrast the Impossible: Region-Constrained Batching for Contrastive User Modeling on a Local Community Platform}

\author{Seungho Han}
\authornote{Both authors contributed equally to this research.}
\email{hawke@daangn.com}
\affiliation{%
  \institution{Danggeun Market Inc. (Karrot)}
  \city{Seoul}
  \country{Korea}
}

\author{Byeongchang Kim}
\authornotemark[1]
\email{ben.kim@daangn.com}
\affiliation{%
  \institution{Danggeun Market Inc. (Karrot)}
  \city{Seoul}
  \country{Korea}
}

\author{Jin Yu}
\email{jin.yu@daangn.com}
\affiliation{%
  \institution{Danggeun Market Inc. (Karrot)}
  \city{Seoul}
  \country{Korea}
}

\renewcommand{\shortauthors}{Seungho Han, Byeongchang Kim, and Jin Yu}

\DeclareRobustCommand\onedot{\futurelet\@let@token\@onedot}
\def\onedot{. }
\def\eg{\emph{e.g}\onedot} \def\Eg{\emph{E.g}\onedot}
\def\ie{\emph{i.e}\onedot} \def\Ie{\emph{I.e}\onedot}
\def\cf{\emph{c.f}\onedot} \def\Cf{\emph{C.f}\onedot}
\def\etc{\emph{etc}\onedot} \def\vs{\emph{vs}\onedot}
\def\wrt{w.r.t\onedot} \def\dof{d.o.f\onedot}
\def\etal{\emph{et al}\onedot}

\begin{abstract}
Contrastive learning is widely used for user modeling in large-scale recommender systems, where standard in-batch negatives implicitly assume universal exposure that any user can be shown any item.
On local community platforms such as Karrot, however, exposure is geographically constrained; many user-item pairs are impossible by design yet still treated as negatives during training, diluting the contrastive learning signal.
We address this impossible negatives problem and propose \textbf{Region-Constrained Batch Sampling (RCBS)}, a simple yet effective batching method that constructs region-homogeneous mini-batches so that users are contrasted primarily against items they could feasibly see.
By replacing impossible negatives with feasible ones, RCBS naturally introduces harder and more informative negatives under realistic exposure constraints.
With offline evaluations and online A/B tests, we show that RCBS consistently improves user representation quality and consequently enhances home feed ranking, retrieval, and display ads ranking.
The resulting user embeddings have been deployed in production across various applications.
\end{abstract}

\ccsdesc[500]{Information systems~Personalization}
\ccsdesc[500]{Information systems~Recommender systems}

\keywords{Recommender Systems, User Representation, Representation Learning, User Modeling}


\maketitle

\section{Introduction}
\begin{figure}[t]
    \centering
    \includegraphics[width=0.9\linewidth]{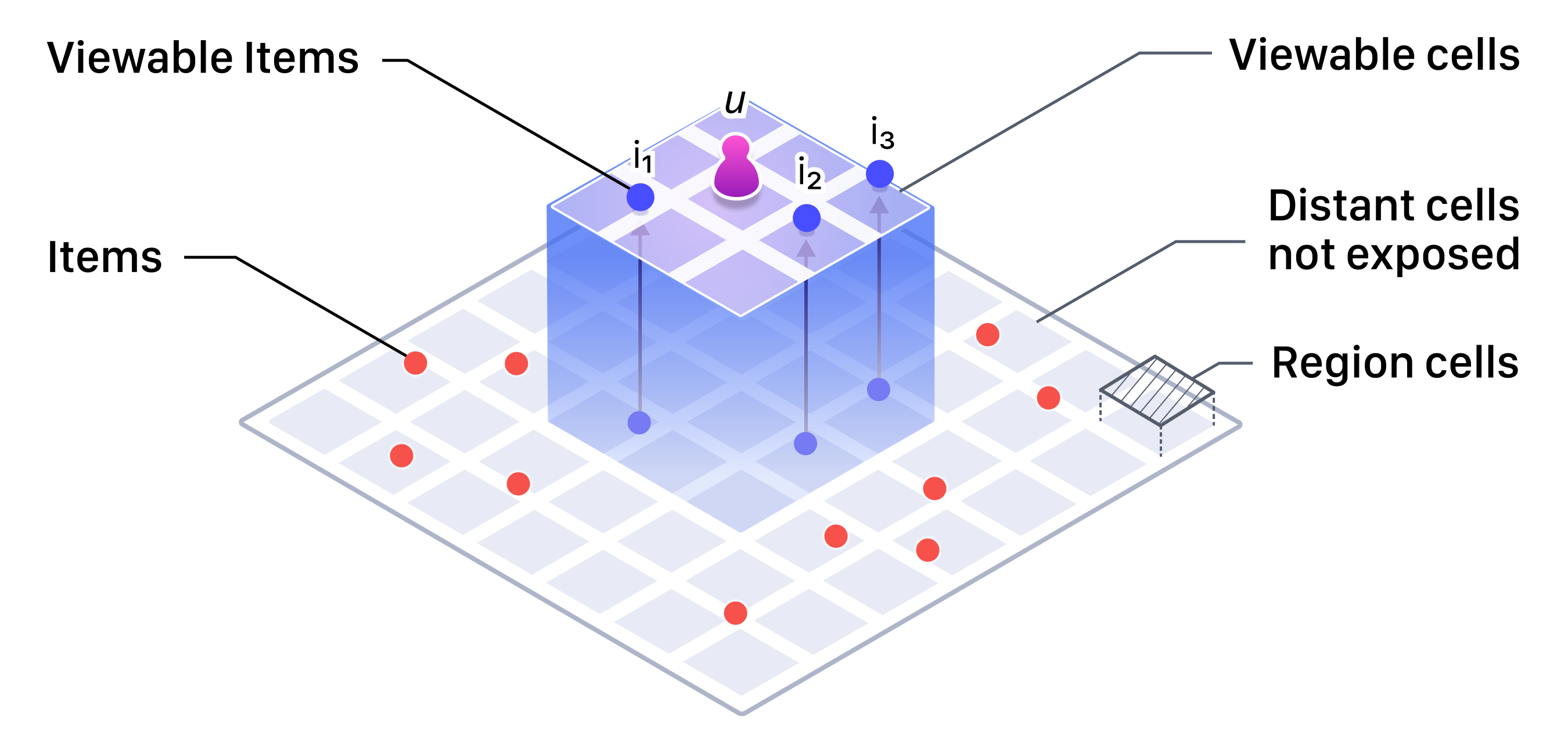}
    \vspace{-10pt}
    \caption{
        Illustration of exposure feasibility on Karrot.
        For user $u$, only items located within the user's viewable cells ($i_1, i_2, i_3$) can appear in the home feed.
        Items outside the viewable cells (red dots) are never exposed to the user.
    }
    \vspace{-10pt}
    \label{fig:karrot_feasibility}
\end{figure}

Contrastive learning is now a standard paradigm for learning user representations in large-scale recommender systems \citep{Pancha:2022:KDD, Chen:2025:RecSys, Zhang:2024:TheWebConf}.
These methods treat observed user-item engagement pairs as positives and other in-batch items as negatives to learn discriminative user representations.
A common, but often implicit, assumption behind random mini-batches is universal exposure: any user could be exposed to any item.
This assumption breaks on \textit{local community platforms}, where exposure is geographically constrained by design.
On Karrot\footnote{\url{https://karrotmarket.com/about/}}, for example, users interact with nearby residents across multiple verticals (local buy \& sell, pre-owned cars, local jobs, social groups, local business, \etc). 
More than 86\% of transactions occur within a 5km radius, and items outside a user’s viewable region are never exposed.
Figure \ref{fig:karrot_feasibility} illustrates the boundaries that define feasibility for exposure, highlighting that only items in the same or adjacent regions appear in a user's home feed.

Under such constraints, random mini-batches inevitably contain negatives that, by platform policy rather than by user preference, the platform does not expose to users.
We distinguish two categories of negatives in geographically constrained contrastive learning:
(i) \textit{feasible negatives}, which a user could have been exposed to but chose not to engage with, and
(ii) \textit{impossible negatives}, which a user could never have been exposed to due to platform's exposure rules.
The former carry a meaningful preference signal, as non-engagement reflects preference; the latter carries little signal since non-engagement is determined by system rather than user.
When impossible negatives dominate a batch, the contrastive signal for preference can weaken and the resulting user representations can become less discriminative.
This issue can extend beyond Karrot to other location-aware services.
For example, restaurant and grocery delivery apps surface only merchants and items within predefined service areas.
Dating apps typically prioritize or filter candidates within distance thresholds.
In all these settings, mini-batches that ignore exposure feasibility risk learning user representations misaligned with user preferences.

To address this, we propose Region-Constrained Batch Sampling (RCBS), a simple yet effective batching method that reduces impossible negatives by constructing region-homogeneous mini-batches.
RCBS leaves the model architecture and loss unchanged and modifies only how batches are formed so that users are contrasted against candidates they could feasibly see.
Our experiments empirically show that feasible negatives act as harder negatives than impossible negatives; correspondingly, a lower impossible negative ratio in a mini-batch correlates with better user-modeling performance (see Section \ref{subsec:results}).
Both offline evaluation and online A/B tests show that the proposed RCBS improves user representation quality and consequently increases retrieval and ranking performance for Karrot's home feed and display ads.
The resulting user embeddings have been deployed in production at Karrot across various applications.

Our contributions are highlighted as follows:
\begin{itemize}
    \item {
        We identify the problem of \textit{impossible negatives} in region-constrained recommendation systems and propose \textit{Region-Constrained Batch Sampling (RCBS)} to remove geographically impossible negatives from in-batch negative sampling.
    }
    \item {
        With offline and online evaluation, we show that RCBS improves user representation quality, and downstream retrieval and ranking performance.
        As a result, user representation learned with RCBS has been deployed in production at Karrot.
    }
\end{itemize}

\section{Related Work}
\label{sec:related_work}

\textbf{Contrastive Learning for User Representation}.
Contrastive learning (CL) is now the de facto standard for learning representations from behavioral logs in recommendation systems \citep{Xie:2021:SIGIR, Wu:2022:SIGIR, Zhang:2024:TheWebConf}.
PinnerFormer \citep{Pancha:2022:KDD} leverages Transformer-based sequence modeling to build user representations at scale.
\citet{Shin:2023:AAAI} shows recommendation models trained with contrastive objective exhibit scaling laws.
Recently, \citet{Chen:2025:RecSys} introduced PinFM, a foundation model trained on Pinterest's massive user activity sequences.
Although our work builds on these advances, to the best of our knowledge we are the first to address \textit{impossible negatives} that arise in region-constrained recommender systems.

\textbf{Hard-negative Mining and Negative Sampling for CL.}
Hard negatives are known to improve performance of contrastive learning \citep{Chuang:2020:NeurIPS, Robinson:2021:ICLR, Shi:2023:TheWebConf}.
In vision-language and metric learning, \citet{Faghri:2018:BMVC} highlights the effectiveness of hard negatives for tightening image-text embeddings,
and \citet{Kalantidis:2020:NeurIPS} propose curriculum-like mixtures to stabilize training.
In NLP, SimCSE \cite{Gao:2021:EMNLP} and RocketQA \cite{Qu:2021:NAACL} mine or synthesize hard negatives for sentence embeddings and dense passage retrieval.
In recommendation, \citet{Yang:2020:TheWebConf} blends multiple negative sources to improve learning signals,
and \citet{Wang:2021:SIGIR} increase the pool of negatives via cross-batch sampling.
In contrast to explicit hard-negative mining, our approach constructs region-constrained mini-batches that remove geographically impossible negatives, thereby naturally introducing feasible negatives as harder negatives.

\textbf{Propensity-Based Debiasing in Recommendations.}
Inverse propensity scoring (IPS) corrects for non-uniform exposure by re-weighting each observed user-item interaction by the inverse of its exposure probability \citep{Schnabel:2016:ICML}.
\citet{Saito:2020:WSDM} extend IPS to implicit feedback under missing-not-at-random assumptions, \citet{Qin:2020:KDD} develop attribute-based propensity estimation for unbiased recommendation, and \citet{Lee:2023:SIGIR} combine IPS with contrastive learning.
All IPS methods require the \textit{positivity assumption}, namely that every user-item pair must have strictly positive exposure probability.
On region-constrained platforms, however, geographic policy creates \textit{structural} positivity violations.
Impossible user-item pairs have zero exposure probability by design, so no interaction exists for IPS to re-weight.
RCBS and IPS are therefore complementary, as RCBS ensures the batch is composed of feasible pairs that carry a meaningful exposure signal, while IPS can correct residual exposure imbalance among them.
We discuss this complementarity in Section~\ref{subsec:rcbs}.

\section{Approach}
\label{sec:approach}

\begin{figure*}[t] \begin{center}
    \centering
    \includegraphics[width=0.8\textwidth]{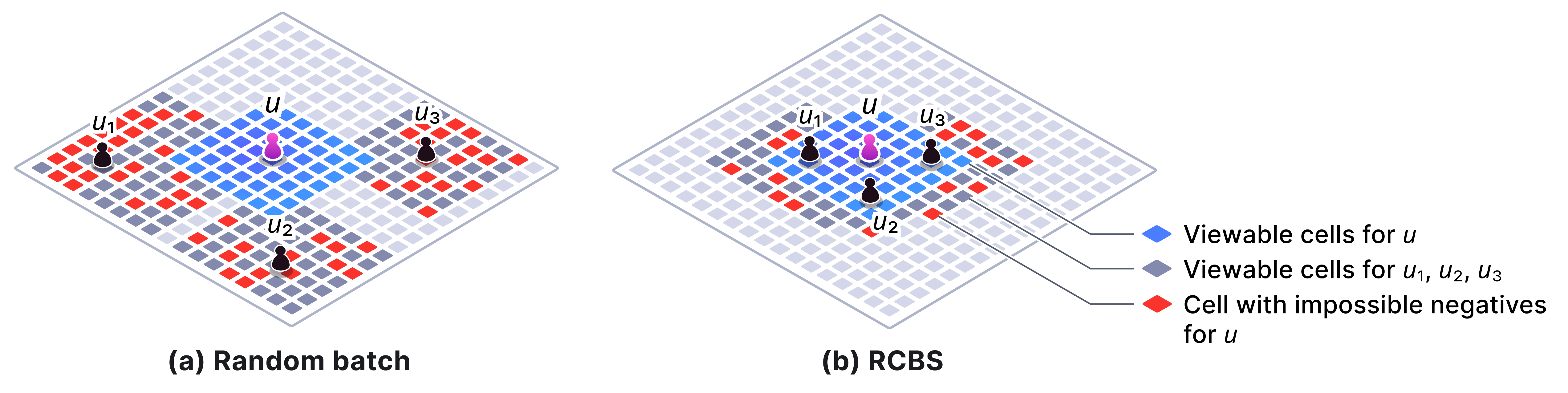}
    \caption{
        Comparison between standard random batching and the proposed Region-Constrained Batch Sampling (RCBS) under region-constrained exposure.
        For a given user $u$, RCBS constructs mini-batches from region-homogeneous users (within blue cells), thereby substantially reducing the presence of impossible negatives (red cells) compared to random batching.
    }
    \label{fig:rcbs}
\end{center} \end{figure*}

We present a batching method for contrastive user modeling under region-constrained exposure conditions.
We first review the standard contrastive learning framework for user modeling (Section \ref{subsec:user_modeling}),
then introduce \emph{Region-Constrained Batch Sampling} (RCBS), where our major contribution lies (Section \ref{subsec:rcbs}).
Figure \ref{fig:rcbs} compares random batching and region-constrained batching under region-constrained exposure conditions.

\subsection{User Modeling with Contrastive Learning}
\label{subsec:user_modeling}

\textbf{Problem Setup}.
We consider a user modeling task where $\mathcal{U}$ denotes the set of users and $\mathcal{I}$ denotes the set of items.
For each user $u \in \mathcal{U}$, we observe a time-ordered sequence of user actions $S_u$ of length $T$:
\begin{align}
    S_u=\{s_{u,1},\dots,s_{u,T}\}, \quad
    s_{u,t} = (a_{u,t}, \tau_{u,t}, i_{u,t}),
\end{align}
where $a_{u,t}$ is the action type (e.g., click, watch, chat), $\tau_{u,t}$ is the timestamp, $i_{u,t} \in \mathcal{I}$ is the item interacted with at step $t$.
Each item $i \in \mathcal{I}$ is represented by its features.
Given a user's action sequence $S_{u,{1:t}}$, our goal is to distinguish the user's next item $i_{u,{t+1}}$ from other candidates.

\textbf{Two-tower Model}.
We use a standard two-tower model \citep{Covington:2016:RecSys} for learning user and item representations.
The user tower $f_{\text{user}}$ encodes the action sequence $S_u$ into a sequence of user representations,
\begin{align}
    \mathbf{H}_u = \{ \mathbf{h}_{u,t} \}_{t=1}^T = f_{\text{user}}(S_u) \in \mathbb{R}^{T \times d},
\end{align}
implemented with a Pre-LN Transformer \citep{Vaswani:2017:NeurIPS, Xiong:2020:ICML}.
The item tower $f_{\text{item}}$ encodes each item into the same embedding space,
\begin{align}
    \mathbf{v}_i = f_{\text{item}}(i) \in \mathbb{R}^d.
\end{align}
We $\ell_2$-normalize both user and item embeddings, and score affinity by the inner product $s(\mathbf{h}_{u,t}, \mathbf{v}_i)=\mathbf{h}_{u,t}^\top\mathbf{v}_i$.

\textbf{Batching and Objective}.
The model is trained using mini-batches of shape $[B, T]$, where $B$ denotes the number of users and $T$ represents the sequence length.
For every position in this sequence, we use the next interacted item as a positive target $\mathbf{v}^+_{b,t}$,
and use all other positive items in the batch as negatives.
We optimize the InfoNCE loss \citep{Oord:2018:arXiv} with temperature $\tau > 0$:
\begin{align}
    \label{eq:infonce}
    \mathcal{L}_{\text{NCE}}
    = -\frac{1}{BT} \sum_{b=1}^B \sum_{t=1}^T
    \log
    \frac{\exp\!\big(s(\mathbf{h}_{b,t}, \mathbf{v}^+_{b,t})/\tau\big)}
         {\sum_{b'=1}^B \sum_{t'=1}^T \exp\!\big(s(\mathbf{h}_{b,t}, \mathbf{v}^+_{b',t'})/\tau\big)}.
\end{align}
This objective encourages the user embedding to align with its 
positive item while separating it from in-batch negatives.

\subsection{Region-Constrained Batch Sampling}
\label{subsec:rcbs}

Standard random batching assumes universal exposure: any item could be exposed to any user.
However, on region-constrained platforms such as Karrot, many user-item pairs are ineligible for exposure, creating \textit{impossible negatives} that misguide the contrastive objective.

\textbf{Exposure Feasibility.}
We partition the geographic space into a discrete set of region cells $r \in \mathcal{R}$.
Each user $u \in \mathcal{U}$ and item $i \in \mathcal{I}$ has location coordinates, which are mapped to specific region cells via a function $r(\cdot)$, yielding $r(u) \in \mathcal{R}$ and $r(i) \in \mathcal{R}$, respectively.
We define exposure feasibility between a user-item pair as:
\begin{align}
    \text{feas}(u, i) = \mathds{1} [\text{dist}(r(u), r(i)) \leq \delta_u],
\end{align}
where $\text{dist}(\cdot, \cdot)$ represents a non-negative, symmetric distance metric between region cells,
and $\delta_u \in \{0, 1, 2, \ldots\}$ denotes a user-specific exposure radius measured in region cells.
In practice, $\delta_u$ is determined by platform policy (\eg user-configured radius), enabling different exposure ranges across users.
When $\delta_u = 0$, exposure requires exact region match, while $\delta_u > 0$ includes items from adjacent regions within the specified threshold.
An \textit{impossible negative} for user $u$ is any negative item $i$ with $\text{feas}(u, i) = 0$.

\textbf{Region-constrained Batching.}
RCBS reduces impossible negatives by forming mini-batches $\mathcal{B}_r$ homogeneous in region:
\begin{align}
    \label{eq:region_batch}
    \mathcal{B}_r = \{(u_b, S_{u_b})\}_{b=1}^M, \quad u_b \sim \text{Uniform}(U_r),
\end{align}
where $U_{r} = \{u \in \mathcal{U} | r(u) = r \}$ is the pool of users located in region $r$.
Since users in $\mathcal{B}_r$ share exposure constraints, most in-batch items are feasible negatives for each user.

Under exact region matching (\ie $\!\delta_u\!=\!0$ for all $u\!\in\!\mathcal{U}$), the expected fraction of impossible negatives within a mini-batch is
\begin{align}
    \label{eq:impossible_ratio}
    \rho = 1 - \sum_{r \in \mathcal{R}}p_r^2,
\end{align}
where $p_r$ is the probability that a user in the batch belongs to region $r$.
For random batching, the probability distribution across regions approximates a uniform distribution, yielding $\rho\!\approx\!1\!-\!\frac{1}{\min(|\mathcal{B}|, |\mathcal{R}|)}$.
This results in a high fraction of impossible negatives that grows with system scale.
On the other hand, RCBS concentrates the probability mass on a single region, achieving $p_r\!\approx\!1$ for the target region and $p_r\!\approx\!0$ for others.
This drives $\rho$ close to zero, independent of batch size $|\mathcal{B}|$ and the number of distinct regions $|\mathcal{R}|$.

In practice, the fraction of impossible negatives $\rho$ for $\mathcal{B}_r$ may not reach exactly zero for two major reasons.
First, varying exposure radii $\delta_u$ across users inevitably introduce some impossible negatives.
For instance, items feasible for users with larger $\delta_u$ may be impossible for users with smaller $\delta_u$.
Second, a user's region $r(u)$ can occasionally change over time (\eg relocation), resulting in historical interactions spanning multiple regions.

After batch construction, we train with the same contrastive objective in Eq.\eqref{eq:infonce}, without modifying the model architecture.

\textbf{Relationship to Propensity-Based Debiasing.}
IPS and RCBS operate at different stages of the training pipeline.
IPS is a \textit{loss-level} correction that re-weights each observed user-item interaction by inverse exposure probability, and therefore requires $p(o{=}1 \mid u,i)>0$ for every pair.
RCBS is a \textit{data-level} correction that changes which items appear in the denominator of Eq.~\eqref{eq:infonce} by forming region-homogeneous mini-batches.
On region-constrained platforms, impossible negatives have $p(o{=}1 \mid u,i)=0$ by policy, so no observed interaction exists for IPS to re-weight.
Under random batching, $\rho_\text{train}{=}0.98$ (Table~\ref{tab:pretrain}), meaning 98\% of in-batch negatives fall into this category.
RCBS reduces $\rho_\text{train}$ to 0.30, ensuring that the majority of negatives are feasible.
Propensity correction can then be applied within a region-homogeneous batch to address residual exposure imbalance among the feasible pairs.

\vspace{-5pt}

\section{Experiments}
\label{sec:experiments}

\subsection{Experimental Setting}
\label{subsec:experimental_setting}
We compare RCBS with standard random batching on a next-action user-modeling pretraining task,
and evaluate the learned user embeddings on three downstream applications: home feed ranking, home feed retrieval, and display ads ranking.

\textbf{Dataset.}
For user-modeling pretraining, we use two years of production logs comprising $\sim$25M users and $\sim$15B actions.
Actions are time-ordered per user, and each user sequence is truncated to at most 1,024 actions.
We split 95\% of users for training, 5\% for evaluation.
For downstream task training, we use service logs that do not overlap with the pretraining period: three weeks for home feed ranking, one week for home feed retrieval, and four weeks for display ads ranking.
Each downstream task employs a time-split evaluation set comprising 24 hours for home feed ranking and display ads ranking, and 6 hours for home feed retrieval.

\textbf{Evaluation Metrics.}
For user-modeling evaluation, we report Recall@10 (R@10) and Recall@100 (R@100) under two distinct evaluation-time batching methods:
(i) \textit{Random-Eval}, in which users are sampled uniformly at random, and
(ii) \textit{RCBS-Eval}, in which users are sampled following RCBS.
We also report the fraction of impossible negatives $\rho_\text{train}$ from Eq.\eqref{eq:impossible_ratio} to quantify the prevalence of impossible negatives within train-time mini-batch.

For downstream tasks, we use standard evaluation metrics for each task.
We use NDCG@10 (N@10) and Ordered-Pair Accuracy (OPA) of click prediction for home-feed ranking,
Recall@10 and Recall@100 for home-feed retrieval, and
ROC-AUC (R-A) and PR-AUC (P-A) for display-ads ranking.
Due to confidentiality, we report relative improvements over the production baselines.

\textbf{User-modeling Pretraining Model.}
We use a standard two-tower architecture, consisting of a Transformer-based user tower and a 3-layer MLP item tower.
For item features, we use embeddings obtained from a pretrained embedding model \citep{Nussbaum:2024:arXiv, Nussbaum:2025:arXiv}, which takes an item’s title, content, metadata and image as input and produces the corresponding item representations.
We train the model with RCBS at two regional granularities supported by our system, and a random batching baseline: (i) \textbf{RCBS-Train (coarse)}, using coarse regional groupings, (ii) \textbf{RCBS-Train (fine)}, using finer regional groupings, and (iii) \textbf{Random-Train}, using random batches.
We use a batch size of 18 and sequence length of 1,024, yielding 18,432 total candidates per mini-batch.
We keep these settings fixed, along with all other hyperparameters and training steps, for fair evaluation.

\textbf{Downstream Task Model.}
We use our production models for downstream tasks.
For home feed ranking and display ads ranking, we use DCN-based models \citep{Wang:2021:TheWebConf} and add the learned user representation as an additional feature.
For home feed retrieval, we use a two-tower model \citep{Covington:2016:RecSys} and inject the user representation into the user tower.
We compute user embeddings daily by running the user tower on a user's recent 1,024 actions and use mean-pooled output as final user representations.

\begin{table}[t]
    \centering
    \small
    \caption{
        Quantitative results on user-modeling pretraining task under different combinations of train-time (rows) and evaluation-time (columns) batching methods.
    }
    \vspace{-8pt}
    \label{tab:pretrain}
    \setlength{\tabcolsep}{6pt}
    \begin{tabular}{lccccc}
        \toprule
                            & \multicolumn{2}{c}{Random-Eval} & \multicolumn{2}{c}{RCBS-Eval (fine)} & \\
                              \cmidrule(lr{1em}){2-3}            \cmidrule(lr{1em}){4-5}
        Method     & R@10 & R@100                    & R@10  & R@100               & $\rho_{\text{train}}$ \\
        \midrule 
        Random-Train        & 0.100    & 0.422     & 0.082     & 0.349     & 0.98 \\
        RCBS-Train (coarse) & 0.125      & 0.456       & 0.112       & 0.403       & 0.79 \\
        RCBS-Train (fine)   & \textbf{0.149}      & \textbf{0.474}       & \textbf{0.139}       & \textbf{0.435}       & \textbf{0.30} \\
        \bottomrule
    \end{tabular}
\end{table}

\begin{table}[t]
    \centering
    \small
    \vspace{-10pt}
    \caption{
        Offline experimental results on downstream tasks.
    }
    \vspace{-8pt}
    \label{tab:downstream-offline-delta}
    \setlength{\tabcolsep}{2.3pt}
    \begin{tabular}{lcccccc}
        \toprule
                        & \multicolumn{2}{c}{Feed ranking}      & \multicolumn{2}{c}{Ads ranking}   & \multicolumn{2}{c}{Feed retrieval} \\
                          \cmidrule(lr){2-3}                    \cmidrule(lr){4-5}                  \cmidrule(lr){6-7}
        Method          & $\Delta$N@10 & $\Delta$OPA            & $\Delta$R-A & $\Delta$P-A         & $\Delta$R@10 & $\Delta$R@100 \\
        \midrule
        Random-Train    & +0.22\% & +0.17\%                     & +0.25\% & +1.21\%	                   & +4.44\% & +2.36\% \\
        RCBS-Train (fine)      & \textbf{+1.18\%} & \textbf{+1.07\%}   & \textbf{+0.53\%} & \textbf{+3.38\%} & \textbf{+7.56\%} & \textbf{+4.61\%} \\
        \bottomrule
    \end{tabular}
    \vspace{-10pt}
\end{table}

\subsection{Results}
\label{subsec:results}

\textbf{User-modeling Experiments.}
Table \ref{tab:pretrain} reports quantitative results on the user-modeling pretraining task.
RCBS-Train consistently outperforms Random-Train on both Random-Eval and RCBS-Eval.
The fraction of impossible negative $\rho_\text{train}$ correlates well with better recall metrics, empirically validating that feasible negatives provide more discriminative training signals.
Notably, the RCBS-Eval consistently yields lower scores compared to Random-Eval,
further supporting the proposition that feasible negatives constitute harder negatives compared to impossible negatives.

\textbf{Offline Experiments.}
Table \ref{tab:downstream-offline-delta} summarizes offline experimental results when user representations are used as additional features in home feed ranking, home feed retrieval, and display-ads ranking.
User embeddings trained with RCBS consistently outperform those learned with random batching across all downstream tasks,
demonstrating enhanced discriminative capability when trained with feasible negatives.
The marginal gains from random-batch user embeddings in home feed ranking highlight the need for RCBS to learn meaningful representations.
The substantially larger gains in home feed retrieval, relative to feed and ads ranking, reflects the structural similarity between the retrieval and pretraining objectives.

\textbf{Online A/B Experiments.}
We select the best user representation, RCBS-Train (fine), based on the offline experiments and conduct A/B tests on home feed recommendation (retrieval + ranking) and display ads ranking.
Table \ref{tab:online-delta} summarizes online experimental results comparing treatment groups (with user representation) against control groups (without user representation).
The consistent improvements across all primary metrics confirm that RCBS improves production metrics, leading to production deployment across multiple applications.

\begin{table}[t]
    \centering
    \small
    \setlength{\tabcolsep}{4pt}
    \caption{
        Online A/B test results when adding our best user embeddings, RCBS-Train (fine), as an additional feature to the production baselines.
        DAV denotes daily active viewers.
    }
    \label{tab:online-delta}
    \begin{tabular}{ccccc}
        \toprule
        \multicolumn{3}{c}{Home feed (ranking + retrieval)}             & \multicolumn{2}{c}{Ads ranking} \\
        \cmidrule(lr){1-3}                                                \cmidrule(lr){4-5}
        $\Delta$Clicks & $\Delta$Impressions & $\Delta$DAV     & $\Delta$eCPM & $\Delta$AdCTR \\
        \midrule
        +10.0\% & +5.12\% & +1.91\%                               & +6.01\% & +7.46\% \\
        \bottomrule
    \end{tabular}
\end{table}

\section{Conclusion}

In this work, we identify and address the impossible negatives problem in region-constrained recommendation systems by introducing RCBS, a batching method that constructs region-homogeneous mini-batches.
We show that feasible negatives provide stronger training signals than impossible negatives, and that RCBS increases the fraction of feasible negatives within each mini-batch.
Experiments show that RCBS improves user representation quality and consequently improves performance on downstream tasks.
Important future directions include exploring adaptive region grouping methods for dynamic feasibility constraints and combining propensity-based correction within region-homogeneous batches to address residual exposure imbalance.


\begin{acks}
We thank Lyn Kim for her help with the figures in this paper, Sang Wook Park and Beomju Kwak for their help with the ads experiments, and Kyuhyun Byun for his help with large-scale inference.
\end{acks}

\section*{Presenter Biography}
Seungho Han is a Tech Lead Manager at Karrot, South Korea’s leading hyper-local community platform.
He has over 6 years of experience in search and recommendation systems.
His research interests include large-scale machine learning with applications in search and recommendation systems.

\balance
\bibliographystyle{ACM-Reference-Format}
\bibliography{sigir_user_modeling}


\begin{thebibliography}{26}


\ifx \showCODEN    \undefined \def \showCODEN     #1{\unskip}     \fi
\ifx \showISBNx    \undefined \def \showISBNx     #1{\unskip}     \fi
\ifx \showISBNxiii \undefined \def \showISBNxiii  #1{\unskip}     \fi
\ifx \showISSN     \undefined \def \showISSN      #1{\unskip}     \fi
\ifx \showLCCN     \undefined \def \showLCCN      #1{\unskip}     \fi
\ifx \shownote     \undefined \def \shownote      #1{#1}          \fi
\ifx \showarticletitle \undefined \def \showarticletitle #1{#1}   \fi
\ifx \showURL      \undefined \def \showURL       {\relax}        \fi
\providecommand\bibfield[2]{#2}
\providecommand\bibinfo[2]{#2}
\providecommand\natexlab[1]{#1}
\providecommand\showeprint[2][]{arXiv:#2}

\bibitem[Chen et~al\mbox{.}(2025)]%
        {Chen:2025:RecSys}
\bibfield{author}{\bibinfo{person}{Xiangyi Chen}, \bibinfo{person}{Kousik Rajesh}, \bibinfo{person}{Matthew Lawhon}, \bibinfo{person}{Zelun Wang}, \bibinfo{person}{Hanyu Li}, \bibinfo{person}{Haomiao Li}, \bibinfo{person}{Saurabh~Vishwas Joshi}, \bibinfo{person}{Pong Eksombatchai}, \bibinfo{person}{Jaewon Yang}, \bibinfo{person}{Yi-Ping Hsu}, \bibinfo{person}{Jiajing Xu}, {and} \bibinfo{person}{Charles Rosenberg}.} \bibinfo{year}{2025}\natexlab{}.
\newblock \showarticletitle{{PinFM: Foundation Model for User Activity Sequences at a Billion-scale Visual Discovery Platform}}. In \bibinfo{booktitle}{\emph{RecSys}}.
\newblock


\bibitem[Chuang et~al\mbox{.}(2020)]%
        {Chuang:2020:NeurIPS}
\bibfield{author}{\bibinfo{person}{Ching-Yao Chuang}, \bibinfo{person}{Joshua~W. Robinson}, \bibinfo{person}{Yen-Chen Lin}, \bibinfo{person}{Antonio Torralba}, {and} \bibinfo{person}{Stefanie Jegelka}.} \bibinfo{year}{2020}\natexlab{}.
\newblock \showarticletitle{{Debiased Contrastive Learning}}. In \bibinfo{booktitle}{\emph{NeurIPS}}.
\newblock


\bibitem[Covington et~al\mbox{.}(2016)]%
        {Covington:2016:RecSys}
\bibfield{author}{\bibinfo{person}{Paul Covington}, \bibinfo{person}{Jay Adams}, {and} \bibinfo{person}{Sargin Emre}.} \bibinfo{year}{2016}\natexlab{}.
\newblock \showarticletitle{{Deep Neural Networks for YouTube Recommendations}}. In \bibinfo{booktitle}{\emph{RecSys}}.
\newblock


\bibitem[Faghri et~al\mbox{.}(2018)]%
        {Faghri:2018:BMVC}
\bibfield{author}{\bibinfo{person}{Fartash Faghri}, \bibinfo{person}{David~J. Fleet}, \bibinfo{person}{Jamie~Ryan Kiros}, {and} \bibinfo{person}{Sanja Fidler}.} \bibinfo{year}{2018}\natexlab{}.
\newblock \showarticletitle{{VSE++: Improving Visual-Semantic Embeddings with Hard Negatives}}. In \bibinfo{booktitle}{\emph{BMVC}}.
\newblock


\bibitem[Gao et~al\mbox{.}(2021)]%
        {Gao:2021:EMNLP}
\bibfield{author}{\bibinfo{person}{Tianyu Gao}, \bibinfo{person}{Xingcheng Yao}, {and} \bibinfo{person}{Danqi Chen}.} \bibinfo{year}{2021}\natexlab{}.
\newblock \showarticletitle{{SimCSE: Simple Contrastive Learning of Sentence Embeddings}}. In \bibinfo{booktitle}{\emph{EMNLP}}.
\newblock


\bibitem[Kalantidis et~al\mbox{.}(2020)]%
        {Kalantidis:2020:NeurIPS}
\bibfield{author}{\bibinfo{person}{Yannis Kalantidis}, \bibinfo{person}{Mert~Bulent Sariyildiz}, \bibinfo{person}{Noe Pion}, \bibinfo{person}{Philippe Weinzaepfel}, {and} \bibinfo{person}{Diane Larlus}.} \bibinfo{year}{2020}\natexlab{}.
\newblock \showarticletitle{{Hard Negative Mixing for Contrastive Learning}}. In \bibinfo{booktitle}{\emph{NeurIPS}}.
\newblock


\bibitem[Lee et~al\mbox{.}(2023)]%
        {Lee:2023:SIGIR}
\bibfield{author}{\bibinfo{person}{Jae-woong Lee}, \bibinfo{person}{Seongmin Park}, \bibinfo{person}{Mincheol Yoon}, {and} \bibinfo{person}{Jongwuk Lee}.} \bibinfo{year}{2023}\natexlab{}.
\newblock \showarticletitle{{uCTRL: Unbiased Contrastive Representation Learning via Alignment and Uniformity for Collaborative Filtering}}. In \bibinfo{booktitle}{\emph{SIGIR}}.
\newblock


\bibitem[Nussbaum and Duderstadt(2025)]%
        {Nussbaum:2025:arXiv}
\bibfield{author}{\bibinfo{person}{Zach Nussbaum} {and} \bibinfo{person}{Brandon Duderstadt}.} \bibinfo{year}{2025}\natexlab{}.
\newblock \showarticletitle{{Training Sparse Mixture Of Experts Text Embedding Models}}. In \bibinfo{booktitle}{\emph{arXiv}}.
\newblock


\bibitem[Nussbaum et~al\mbox{.}(2024)]%
        {Nussbaum:2024:arXiv}
\bibfield{author}{\bibinfo{person}{Zach Nussbaum}, \bibinfo{person}{Brandon Duderstadt}, {and} \bibinfo{person}{Andrly Mulyar}.} \bibinfo{year}{2024}\natexlab{}.
\newblock \showarticletitle{{Nomic Embed Vision: Expanding the Latent Space}}. In \bibinfo{booktitle}{\emph{arXiv}}.
\newblock


\bibitem[Oord et~al\mbox{.}(2018)]%
        {Oord:2018:arXiv}
\bibfield{author}{\bibinfo{person}{Aaron van~den Oord}, \bibinfo{person}{Yazhe Li}, {and} \bibinfo{person}{Oriol Vinyals}.} \bibinfo{year}{2018}\natexlab{}.
\newblock \showarticletitle{{Representation Learning with Contrastive Predictive Coding}}. In \bibinfo{booktitle}{\emph{arXiv}}.
\newblock


\bibitem[Pancha et~al\mbox{.}(2022)]%
        {Pancha:2022:KDD}
\bibfield{author}{\bibinfo{person}{Nikil Pancha}, \bibinfo{person}{Andrew Zhai}, \bibinfo{person}{Jure Leskovec}, {and} \bibinfo{person}{Charles Rosenberg}.} \bibinfo{year}{2022}\natexlab{}.
\newblock \showarticletitle{{PinnerFormer: Sequence Modeling for User Representation at Pinterest}}. In \bibinfo{booktitle}{\emph{KDD}}.
\newblock


\bibitem[Qin et~al\mbox{.}(2020)]%
        {Qin:2020:KDD}
\bibfield{author}{\bibinfo{person}{Zhen Qin}, \bibinfo{person}{Suming~Jeremiah Chen}, \bibinfo{person}{Don Metzler}, \bibinfo{person}{Yongwoo Noh}, \bibinfo{person}{Jingzheng Qin}, {and} \bibinfo{person}{Xuanhui Wang}.} \bibinfo{year}{2020}\natexlab{}.
\newblock \showarticletitle{{Attribute-based Propensity for Unbiased Learning in Recommender Systems: Algorithm and Case Studies}}. In \bibinfo{booktitle}{\emph{KDD}}.
\newblock


\bibitem[Qu et~al\mbox{.}(2021)]%
        {Qu:2021:NAACL}
\bibfield{author}{\bibinfo{person}{Yingqi Qu}, \bibinfo{person}{Yuchen Ding}, \bibinfo{person}{Jing Liu}, \bibinfo{person}{Kai Liu}, \bibinfo{person}{Ruiyang Ren}, \bibinfo{person}{Wayne~Xin Zhao}, \bibinfo{person}{Daxiang Dong}, \bibinfo{person}{Hua Wu}, {and} \bibinfo{person}{Haifeng Wang}.} \bibinfo{year}{2021}\natexlab{}.
\newblock \showarticletitle{{RocketQA: An Optimized Training Approach to Dense Passage Retrieval for Open-Domain Question Answering}}. In \bibinfo{booktitle}{\emph{NAACL}}.
\newblock


\bibitem[Robinson et~al\mbox{.}(2021)]%
        {Robinson:2021:ICLR}
\bibfield{author}{\bibinfo{person}{Joshua Robinson}, \bibinfo{person}{Ching-Yao Cuang}, \bibinfo{person}{Suvrit Sra}, {and} \bibinfo{person}{Stefanie Jegelka}.} \bibinfo{year}{2021}\natexlab{}.
\newblock \showarticletitle{{Contrastive Learning with Hard Negative Samples}}. In \bibinfo{booktitle}{\emph{ICLR}}.
\newblock


\bibitem[Saito et~al\mbox{.}(2020)]%
        {Saito:2020:WSDM}
\bibfield{author}{\bibinfo{person}{Yuta Saito}, \bibinfo{person}{Suguru Yaginuma}, \bibinfo{person}{Yuta Nishino}, \bibinfo{person}{Hayato Sakata}, {and} \bibinfo{person}{Kazuhide Nakata}.} \bibinfo{year}{2020}\natexlab{}.
\newblock \showarticletitle{{Unbiased Recommender Learning from Missing-Not-At-Random Implicit Feedback}}. In \bibinfo{booktitle}{\emph{WSDM}}.
\newblock


\bibitem[Schnabel et~al\mbox{.}(2016)]%
        {Schnabel:2016:ICML}
\bibfield{author}{\bibinfo{person}{Tobias Schnabel}, \bibinfo{person}{Adith Swaminathan}, \bibinfo{person}{Ashudeep Singh}, \bibinfo{person}{Navin Chandak}, {and} \bibinfo{person}{Thorsten Joachims}.} \bibinfo{year}{2016}\natexlab{}.
\newblock \showarticletitle{{Recommendations as Treatments: Debiasing Learning and Evaluation}}. In \bibinfo{booktitle}{\emph{ICML}}.
\newblock


\bibitem[Shi et~al\mbox{.}(2023)]%
        {Shi:2023:TheWebConf}
\bibfield{author}{\bibinfo{person}{Wentao Shi}, \bibinfo{person}{Jiawei Chen}, \bibinfo{person}{Fuli Feng}, \bibinfo{person}{Jizhi Zhang}, \bibinfo{person}{Junkang Wu}, \bibinfo{person}{Chongming Gao}, {and} \bibinfo{person}{Xiangnan He}.} \bibinfo{year}{2023}\natexlab{}.
\newblock \showarticletitle{{On the Theories Behind Hard Negative Sampling for Recommendation}}. In \bibinfo{booktitle}{\emph{TheWebConf}}.
\newblock


\bibitem[Shin et~al\mbox{.}(2023)]%
        {Shin:2023:AAAI}
\bibfield{author}{\bibinfo{person}{Kyuyong Shin}, \bibinfo{person}{Hanock Kwak}, \bibinfo{person}{Su~Young Kim}, \bibinfo{person}{Max~Nihlen Ramstrom}, \bibinfo{person}{Jisu Jeong}, \bibinfo{person}{Jung-Woo Ha}, {and} \bibinfo{person}{Kyung-Min Kim}.} \bibinfo{year}{2023}\natexlab{}.
\newblock \showarticletitle{{Scaling Law for Recommendation Models: Towards General-purpose User Representations}}. In \bibinfo{booktitle}{\emph{AAAI}}.
\newblock


\bibitem[Vaswani et~al\mbox{.}(2017)]%
        {Vaswani:2017:NeurIPS}
\bibfield{author}{\bibinfo{person}{Ashish Vaswani}, \bibinfo{person}{Noam Shazeer}, \bibinfo{person}{Niki Parmar}, \bibinfo{person}{Jakob Uszkoreit}, \bibinfo{person}{Llion Jones}, \bibinfo{person}{Aidan~N. Gomez}, \bibinfo{person}{Lukasz Kaiser}, {and} \bibinfo{person}{Illia Polosukhin}.} \bibinfo{year}{2017}\natexlab{}.
\newblock \showarticletitle{{Attention is All You Need}}. In \bibinfo{booktitle}{\emph{NeurIPS}}.
\newblock


\bibitem[Wang et~al\mbox{.}(2021b)]%
        {Wang:2021:SIGIR}
\bibfield{author}{\bibinfo{person}{Jinpeng Wang}, \bibinfo{person}{Jieming Zhu}, {and} \bibinfo{person}{Xiuqiang He}.} \bibinfo{year}{2021}\natexlab{b}.
\newblock \showarticletitle{{Cross-Batch Negative Sampling for Training Two-Tower Recommenders}}. In \bibinfo{booktitle}{\emph{SIGIR}}.
\newblock


\bibitem[Wang et~al\mbox{.}(2021a)]%
        {Wang:2021:TheWebConf}
\bibfield{author}{\bibinfo{person}{Ruoxi Wang}, \bibinfo{person}{Rakesh Shivanna}, \bibinfo{person}{Derek~Z. Cheng}, \bibinfo{person}{Sagar Jain}, \bibinfo{person}{Dong Lin}, \bibinfo{person}{Lichan Hong}, {and} \bibinfo{person}{Ed~H. Chi}.} \bibinfo{year}{2021}\natexlab{a}.
\newblock \showarticletitle{{DCN V2: Improved Deep \& Cross Network and Practical Lessons for Web-scale Learning to Rank Systems}}. In \bibinfo{booktitle}{\emph{TheWebConf}}.
\newblock


\bibitem[Wu et~al\mbox{.}(2022)]%
        {Wu:2022:SIGIR}
\bibfield{author}{\bibinfo{person}{Chuhan Wu}, \bibinfo{person}{Fangzhao Wu}, \bibinfo{person}{Yang Yu}, \bibinfo{person}{Tao Qi}, \bibinfo{person}{Yongfeng Huang}, {and} \bibinfo{person}{Xing Xie}.} \bibinfo{year}{2022}\natexlab{}.
\newblock \showarticletitle{{UserBERT: Contrastive User Model Pre-training}}. In \bibinfo{booktitle}{\emph{SIGIR}}.
\newblock


\bibitem[Xie et~al\mbox{.}(2021)]%
        {Xie:2021:SIGIR}
\bibfield{author}{\bibinfo{person}{Xu Xie}, \bibinfo{person}{Fei Sun}, \bibinfo{person}{Zhaoyang Liu}, \bibinfo{person}{Shiwen Wu}, \bibinfo{person}{Jinyang Gao}, \bibinfo{person}{Bolin Ding}, {and} \bibinfo{person}{Bin Cui}.} \bibinfo{year}{2021}\natexlab{}.
\newblock \showarticletitle{{Contrastive Learning for Sequential Recommendation}}. In \bibinfo{booktitle}{\emph{SIGIR}}.
\newblock


\bibitem[Xiong et~al\mbox{.}(2020)]%
        {Xiong:2020:ICML}
\bibfield{author}{\bibinfo{person}{Ruibin Xiong}, \bibinfo{person}{Yunchang Yang}, \bibinfo{person}{Di He}, \bibinfo{person}{Kai Zheng}, \bibinfo{person}{Shuxin Zheng}, \bibinfo{person}{Chen Xing}, \bibinfo{person}{Huishuai Zhang}, \bibinfo{person}{Yanyan Lan}, \bibinfo{person}{Liwei Wang}, {and} \bibinfo{person}{Tie-Yan Liu}.} \bibinfo{year}{2020}\natexlab{}.
\newblock \showarticletitle{{On Layer Normalization in the Transformer Architecture}}. In \bibinfo{booktitle}{\emph{ICML}}.
\newblock


\bibitem[Yang et~al\mbox{.}(2020)]%
        {Yang:2020:TheWebConf}
\bibfield{author}{\bibinfo{person}{Ji Yang}, \bibinfo{person}{Xinyang Yi}, \bibinfo{person}{Derek~Zhiyuan Cheng}, \bibinfo{person}{Lichang Hong}, \bibinfo{person}{Yang Li}, \bibinfo{person}{Simon Wang}, \bibinfo{person}{Taibai Xu}, {and} \bibinfo{person}{Ed~H. Chi}.} \bibinfo{year}{2020}\natexlab{}.
\newblock \showarticletitle{{Mixed Negative Sampling for Learning Two-tower Neural Networks in Recommendations}}. In \bibinfo{booktitle}{\emph{TheWebConf}}.
\newblock


\bibitem[Zhang et~al\mbox{.}(2024)]%
        {Zhang:2024:TheWebConf}
\bibfield{author}{\bibinfo{person}{Wei Zhang}, \bibinfo{person}{Dai Li}, \bibinfo{person}{Chen Liang}, \bibinfo{person}{Fang Zhou}, \bibinfo{person}{Zhongke Zhang}, \bibinfo{person}{Xuewei Wang}, \bibinfo{person}{Ru Li}, \bibinfo{person}{Yi Zhou}, \bibinfo{person}{Yaning Huang}, \bibinfo{person}{Dong Liang}, \bibinfo{person}{Kai Wang}, \bibinfo{person}{Zhangyuan Wang}, \bibinfo{person}{Zhengxing Chen}, \bibinfo{person}{Fenggang Wu}, \bibinfo{person}{Minghai Chen}, \bibinfo{person}{Huayu Li}, \bibinfo{person}{Yunnan Wu}, \bibinfo{person}{Zhan Shu}, \bibinfo{person}{Mindi Yuan}, {and} \bibinfo{person}{Sri Reddy}.} \bibinfo{year}{2024}\natexlab{}.
\newblock \showarticletitle{{Scaling User Modeling: Large-scale Online User Representations for Ads Personalization in Meta}}. In \bibinfo{booktitle}{\emph{TheWebConf}}.
\newblock


\end{thebibliography}



\end{document}